\documentclass[prl,twocolumn,floatfix,showpacs]{revtex4}
\usepackage{graphicx}
\usepackage{amssymb}
\newcommand{\be}{\begin{equation}}
\newcommand{\ee}{\end{equation}}
\newcommand{\bea}{\begin{eqnarray}}
\newcommand{\eea}{\end{eqnarray}}
\newcommand{\non}{\nonumber}
\def\rnum#1{\expandafter{\romannumeral #1}} 
\def\Rnum#1{\uppercase\expandafter{\romannumeral #1}} 

\begin{document}
\title{Exact Yrast Spectra of Cold Atoms on a Ring}

\author{Eriko Kaminishi}\email{g0940615@edu.cc.ocha.ac.jp}
\author{Rina Kanamoto}\email{kanamoto.rina@ocha.ac.jp}
\author{Jun Sato}\email{jsato@sofia.phys.ocha.ac.jp}
\author{Tetsuo Deguchi}\email{deguchi@phys.ocha.ac.jp}
\affiliation{Department of Physics, 
Graduate School of Humanities and Sciences, 
Ochanomizu University, 2-1-1
Ohtsuka, Bunkyo-ku, Tokyo 112-8610, Japan}

\date{\today}

% ___________________________________________________________________________________________

\begin{abstract}
We propose a methodology to construct excited states with a fixed angular momentum, namely, ``yrast excited states'' of finite-size one-dimensional bosonic systems with periodic boundary conditions. 
The excitation energies such as the first yrast excited energy 
are calculated through the system-size asymptotic expansion 
and expressed analytically by dressed energy. 
 Interestingly, they are grouped into sets of almost degenerate energy levels. 
The low-lying excitation spectrum near the yrast state 
is consistent with the $ U(1)$ conformal field theories 
if the total angular momentum is given by an integral multiple of particle number; i.e., if the system is supercurrent. 
\end{abstract} 
\pacs{03.75.Hh,03.75.Lm}
\maketitle
% ___________________________________________________________________________________________

Ultracold quantum gases have attracted great interest due to their potential application to the testing of the quantum fluctuations in many-body systems~\cite{BDZ08}. 
Nontrivial correlations in one-dimensional exactly solvable quantum many-body systems have been theoretically studied in 
the framework of the Bethe ansatz and conformal field theories (CFT) with finite-size scaling~\cite{Korepin, Kawakami}. 
While these systems are known to reveal large quantum fluctuations, 
there had been no ideal experimental testing 
ground until various control techniques in 
ultracold gas, such as changing the effective spatial dimension, 
sign and strength of interaction~\cite{Feshbach}, became available.

A periodic one-dimensional (1D) bosonic system, for example, Lieb-Liniger (LL) model~\cite{Lieb-Liniger}, is simple 
mathematical model with various applications to the study of many-body phenomena in a wide range of interactions~\cite{TG}. 
The periodic geometry has been experimentally realized as a circular wave guide or elliptic trap~\cite{1Dex,NIST}. 
In periodic 1D systems it is possible to study rotational properties by 
giving finite angular momentum 
to the system. Because the LL model has translation symmetry, 
it is a useful tool for studying the energy spectrum 
within the Hilbert subspace of a fixed angular momentum. 
This concept, originally called the yrast problem 
in nuclear physics~\cite{BP99}, 
is known to provide an insight into ensembles under rotation, 
the generation of topological defects~\cite{BP99,BR99, IT80, Viefers,KCU10}, and the study of quantum Hall phase transitions~\cite{GSC10, Haldane}. 
In particular, two-dimensional yrast problem has been well studied in the context of vortex creation in a harmonic trap. 
The yrast state with the angular momentum of non-integral multiple of $\hbar N$ is regarded as an off-centered vortex state when it is viewed 
from the Gross-Pitaevskii mean-field picture, while that of $\pm \hbar N$ is the centered single vortex state~\cite{BR99}. 
This approach would also give some important clues to understanding the fundamental properties of superfluidity~\cite{LA67,Cooper,Brand}.

In this Letter we construct low-lying excited states for the LL model with a fixed angular momentum. 
The yrast states in the  LL model are known to correspond to the lowest-energy hole excitations, the set of which 
are called the type-II excitation branch in the thermodynamic limit~\cite{Lieb63}. 
However, we note that the concept of the yrast problem is defined only in a finite-size system. 
We define, in this letter, the excitation energies from the yrast state as ``excited yrast energies". 
We analytically express the lowest-order yrast excitation energies in terms of 
dressed energies through the asymptotic expansion with respect to the system size. 
We evaluate these excitation energies by obtaining the particle quasimomenta 
as an asymptotic solution of the Bethe ansatz equations, and by defining the hole quasimomenta. 
Such excitations play a fundamental role in the spectral analysis related to 
scattering experiments~\cite{Stewart} when the system under probe has finite angular momentum.

% ___________________________________________________________________________________________

The LL Hamiltonian for $N$ bosons interacting with $\delta$-function potential 
in 1D is given by
\be 
{\cal H}_{\rm LL} = - {\frac {\hbar^2} {2m}} 
\sum_{j=1}^N {\frac {\partial^2} {{\partial x_j}^2}} 
+ g_{\rm 1D} \sum_{j < k} \delta(x_j - x_k) \, . \label{eq:LL} 
\ee
Here $g_{\rm 1D}$ is the 1D coupling constant~\cite{Ol98}. 
We consider a circular geometry with the radius $R$ by imposing periodic boundary conditions. 
The coordinates $x_j$ are given as $x_j=R \theta_j$ with $\theta_j$ being the azimuthal angle $0 \le \theta_j < 2\pi$, 
and $\ell= 2 \pi R$ the circumference of the ring joining these coordinates. 

% ___________________________________________________________________________________________

We first discuss the Bethe ansatz theory which provides {\it exact} eigensolutions to the Hamiltonian~(\ref{eq:LL}). 
The set of quasimomenta $k_j$'s for $j \in \{1, 2, \ldots, N\}$,  
satisfies the Bethe ansatz equations (BAE)
\be 
\ell k_j = 2\pi I_j - 2 \sum_{\alpha \ne j}^N \arctan\left(\frac{k_j- k_{\alpha}}{u}\right)
\, .  \label{eq:BAE}  
\ee
Here $u= m g_{\rm 1D}/\hbar^2$ is the coupling constant with the dimension of wavenumber. 
Each eigensolution of the Hamiltonian~(\ref{eq:LL}) is specified by a set of quantum numbers $I_j$, 
which are integers for odd $N$, and half-integers for even $N$. 
The ground-state solution, for instance, corresponds to the set of integers 
$I_j^{\rm (g)}= j - (N+1)/2$ for $j\in \{1, 2, \ldots, N\}$. 
It is convenient to define the maximum and minimum quantum numbers (fermi points) for the ground state as 
$I^{\pm}\equiv \pm (N-1)/2$. 
By solving the BAE for a given set of $I_j$, we obtain the set of quasimomenta 
$k_j$. 
Then the eigenvalue of the Hamiltonian~(\ref{eq:LL}) is obtained by summing up $k_j^2$~\cite{Lieb-Liniger}. 
The total angular momentum is given by $L= R \hbar \sum_{j=1}^{N} k_j$, or, 
according to the BAE, 
%~(\ref{eq:BAE}), 
alternatively by 
\be \label{eq:L}
L = \hbar \sum_{j=1}^{N} I_j \, . 
\ee
Hereafter we shall often denote $L/\hbar$ by $L$ for simplicity. 
In the strong coupling Tonks-Girardeau (TG) limit $u\to \infty$, the second term on the right-hand side of 
Eq.~(\ref{eq:BAE}), approaches zero, and the solutions are, therefore, given by $k_j \ell=2\pi I_j$.

% ___________________________________________________________________________________________

In order to evaluate quasimomenta $k_j$ of an arbitrary eigensolution, we introduce a function 
\be 
Z_{\ell}(k) = {\frac k {2\pi}} + {\frac 1 {2 \pi \ell}} 
\sum_{j=1}^{N} 2 \arctan \left(\frac{k-k_j}{u}\right)  \, . \label{eq:Z}
\ee 
In terms of $Z_{\ell}(k)$, the set of BAE~(\ref{eq:BAE}) is simplified to 
\be\label{eq:Zell}
Z_{\ell}(k_j) = \frac{I_j}{\ell} \quad 
\mbox{for} \, \, j \in \{1, 2, \ldots, N\} . 
%(j=1, 2, \ldots, N). 
\ee
When $N$ and $\ell$ are finite but large, the summation of $k_j$ is approximated to an integral including a finite-size correction 
with the use of the Euler-Maclaurin expansion for large $\ell$. 
We investigate the approximate BAE solutions for the ground state and excited state of a pair of particle-hole excitation as follows. 

%%%%%%%  ground-state k %%%%%%%%%%%%%%%%%%%%%%%%%%%%%%%%%%%%%%%%%%%

It is easy to show that the ground-state root density $\rho(k)$ 
satisfies the integral equation 
\be \label{eq:rho}
\rho(k) = {\frac 1 {2 \pi}} + \int_{-Q_0}^{Q_0} {\frac 1 {2 \pi}} 
K(k-q) \rho(q) dq + O(1/\ell^2),  \label{eq:rho}   
\ee
where $K(k)=  2 u/(k^2+ u^2)$. 
We have defined  $\pm Q_0$ by $Z_{\ell}(\pm Q_0) = (I^{\pm}\pm 1/2) / \ell$. 
They are related to the particle density by 
$N/\ell = \int_{-Q_0}^{Q_0} \rho(k)dk$. 
Approximating $Z_{\ell}(k)$ we define a ground-state counting function 
$Z(k)$ as an integral of the root density 
$Z(k)=\int^{k}_{-Q_0} \rho(q) dq + Z(-Q_0)$, where 
$\rho(q)$ denotes a solution to Eq.~(\ref{eq:rho}). 
With $Z(k)$ as a function of $k$, 
we have a mapping between $k_j$ and $Z(k_j)=I_j/\ell$ via Eq.~(\ref{eq:Zell}). 
%The possible values of $k_j$'s are given from an inverse function of $Z$; 
We evaluate $k_j$'s by the inverse function of $Z$; 
i.e., $k_j = Z^{-1}(I_j/\ell)$. 
The difference between the (particle) quasimomenta, evaluated by the inverse of $Z$, and exact solutions, as direct solutions of the BAE, 
is of the order $O(1/\ell^2)$. Here, we derive  $Z(k)$ from  
$Z_{\ell}(k)$ replacing the sum with the integral, 
which gives corrections of the order $O(1/\ell^2)$. 
We thus denote the ground-state quasimomenta 
$k_j =Z^{-1}(I_j^{(g)}/\ell)$ by $k[I_j^{(g)}]$. 

%%%%%%%  excited-state \tilde{k} %%%%%%%%%%%%%%%%%%%%%%%%%%%%%%%%%%%%%%%%%%

We derive a particle-hole excited state from the ground state 
making a hole at $I_h$ and a particle at $I_p$. 
We denote by $\tilde{k}_j$  
the quasimomenta satisfying the BAE
%~(\ref{eq:BAE}) 
with $I_j=I_j^{(g)}$ 
for $1 \le j \le N$ and $j \ne j_h$ ($j_h=I_h + (N+1)/2$),  
and by ${\tilde k}_p$ that of $I_p$. 
We define ${\tilde Z}_{\ell}(k)$ 
replacing $k_j$'s in Eq. (\ref{eq:Z}) 
with ${\tilde k}_j$'s and ${\tilde k}_p$. 
We define the hole quasimomentum ${\tilde k}_h$ by ${\tilde Z}_{\ell}({\tilde k}_h)=I_h/\ell$. In the ground state we define 
${k}_h$ by ${Z}_{\ell}({k}_h)=I_h/\ell$. 
The root density $\rho^{p, h}_{\ell}(\tilde{k})=d{\tilde Z}_{\ell}(\tilde{k})/d\tilde{k}$ satisfies  
\bea 
\rho^{p, h}_{\ell} (\tilde{k}) &=& {\frac 1 {2 \pi}} 
 + \int_{Q^{-}}^{Q^{+}} {\frac {dq} {2 \pi}} K(\tilde{k}-q) \rho_{\ell}^{p, h} (q)   \non \\ 
&& - {\frac 1 {\ell}}\left[ \delta(\tilde{k}-{\tilde k}_h)- \delta(\tilde{k}-{\tilde k}_p)\right] \non \\ 
& &  
+ {\frac 1 {48 \pi \ell^2}} \left(  {\frac {K^{'}(\tilde{k}-Q^{+})} 
{\rho_{\ell}^{p, h}(Q^{+})}}+ {\frac {K^{'}(\tilde{k}-Q^{-})} {\rho_{\ell}^{p, h}(Q^{-})}} \right) \non \\ 
& & 
+ O(\ell^{-3}) .  \label{eq:rho-h} 
\eea
Here $Q^{\pm}$ are related to $N$ and $L$ by   
$\ell \int_{Q^{-}}^{Q^{+}} \rho_{\ell}^{p, h}(\tilde{k})d\tilde{k}= N$ and 
$ R \ell \int_{Q^{-}}^{Q^{+}} \tilde{k} \rho_{\ell}^{p, h} (\tilde{k})d\tilde{k}=L$, 
respectively.

%%%%%%%  dressed energy %%%%%%%%%%%%%%%%%%%%%%%%%%%%%%%%%%%%%%%%%%

Shifting the Hamiltonian as 
${\cal H}_{L, \, N} = {\cal H}_{\rm LL} - \mu N - \Omega L$,   
we define the dressed energy~\cite{Korepin}  by 
\be \label{eq:epsilon}
\epsilon(k) = \epsilon_0(k) + 
\int_{Q^{-}}^{Q^{+}} {\frac 1 {2\pi}} 
K(k-q) \epsilon(q) dq  \, ,
\ee
where $\epsilon_0 (k)= (\hbar k)^2 /(2m) - \mu - \Omega R\hbar k $. 
We determine $\mu$ and $\Omega$ so that 
the dressed energy vanishes at $Q^{\pm}$: 
\be
\epsilon(Q^{+}) = \epsilon(Q^{-})=0 . \label{eq:zero}
\ee
In the asymptotic expansion with respect to $\ell$, 
we obtain by Eq.~(\ref{eq:rho-h}) 
the particle-hole excitation energy $E_{L, N}^{p, h}$ as  
\be 
E_{L, N}^{p, h } = \ell \int^{Q^{+}}_{Q^{-}} \epsilon(k) dk 
+ \epsilon({\tilde k}_p)  - \epsilon({\tilde k}_h) 
- {\frac {\pi v_F} {6 \ell}} +O(\ell^{-2}) . 
\label{eq:finite-energy} 
\ee
The exact result generalizes both the type I and II excitation branches~\cite{Lieb63}, including the finite-size correction of the order $O(1/\ell)$.

%%%%%%%%%%%%%%%%%%%%% figure %%%%%%%%%%%%%%%%%%%%%%%%%%%
\begin{figure}[t]
\includegraphics[width=1.0\columnwidth]{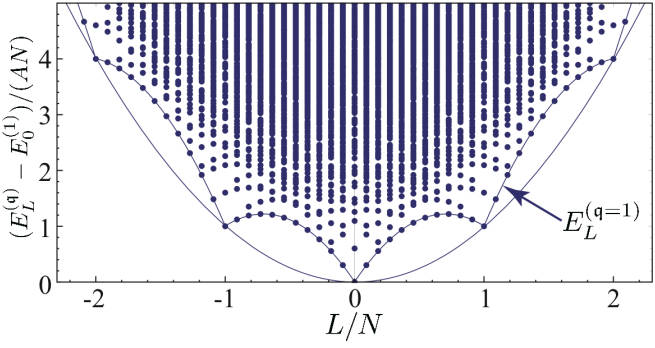}
\caption{(Color online) Eigenvalues $E_L^{(\mathfrak{q})}$ 
of LL Hamiltonian measured from the ground-state energy $E_0^{(1)}$ 
for $u=1.0$ ($N=11$) by solving the BAE. }
\label{fig1}
\end{figure}
%%%%%%%%%%%%%%%%%%%%%%%%%%%%%%%%%%%%%%%%%%%%%%%%%%%

%%%%%%%  1st yrast excited state (exact expression) %%%%%%%%%%%%%%%%%%%%%%%%%%%%%%%%%%%%%%%%%%

For a given angular momentum $L$ with fixed particle number $N$    
we can construct the $\mathfrak{q}$-th  
eigenvalue $E^{(\mathfrak{q})}_{L,N}$ systematically 
for $\mathfrak{q}\in \mathbb{N}$.  
We denote the particle-hole quantum numbers 
 as $I_{p,h}^{(\mathfrak{q})}$ for the $\mathfrak{q}$-th state. 
It is sufficient to consider only a domain $0 \le L < N$ due 
to the Galilean invariance. 
We often simplify $E^{(\mathfrak{q})}_{L,N}$ as $E^{(\mathfrak{q})}_{L}$. 
In Fig.~\ref{fig1} eigenvalues $E^{(\mathfrak{q})}_{L}$ 
are plotted, where the lowest line 
shows the yrast spectrum $\mathfrak{q}=1$ as a function of $L/N$.

\begin{figure}[b]
\includegraphics[width=1.0\columnwidth]{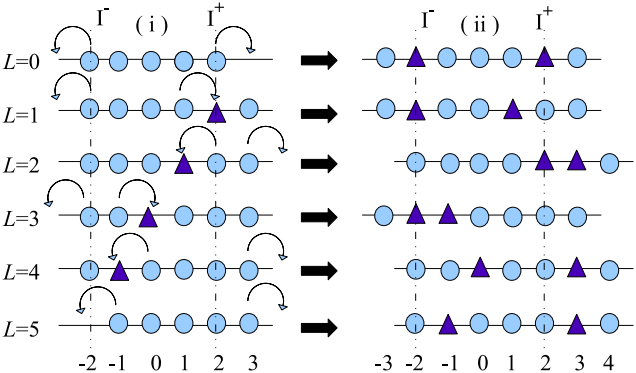}
\caption{(Color online) The scheme for constructing the (i) yrast ground $(\mathfrak{q}=1)$ and (ii) first excited $(\mathfrak{q}=2)$ 
states for $N=5$. The integers denote the quantum numbers $I_j$'s. 
(i) A single hole is present at $I_h^{(1)}$. 
(ii) There are two types of excitations, where 
a particle and a hole move to the left ($L=0,1,3$), 
and to the right ($L=2,4,5$). 
The change in the energy associated with the left and right excitations is denoted as $\Delta E_L^{-}$ and $\Delta E_L^+$, respectively.}
\label{fig2} 
\end{figure}

Let us construct the yrast state $\mathfrak{q}=1$ (type II branch found by Lieb in the thermodynamics limit~\cite{Lieb63}). 
It is derived from the ground state of $L=0$ 
by making a hole at $I_h^{(1)}= I^{+} +1 - L$ and a particle 
at $I_p^{(1)}=I^{+}+1$, as shown in Fig.~\ref{fig2} (i).  
Here we remark that $I_p^{(1)}-I_h^{(1)}=L$. 
We evaluate the quasimomenta by  
$\tilde{k}^{(1)}_j \equiv \tilde{k}[I_j^{(1)}]$.

%%%%%%%%%%%%%%%%%%%%% figure %%%%%%%%%%%%%%%%%%%%%%%%%%%
%%%%%%%%%%%%%%%%%%%%%%%%%%%%%%%%%%%%%%%%%%%%%%%%%%%

\begin{figure}[t]
\includegraphics[width=1.0\columnwidth]{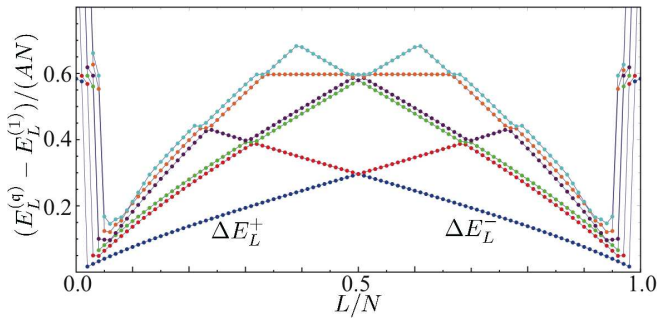}
\caption{(Color online) Eigenvalues $E_L^{(\mathfrak{q})}$ 
of LL Hamiltonian measured from the yrast state $E_L^{(1)}$ for 
$u=1.0$  ($N=100$) through solutions of the BAE, where 
$ E_L^{(\mathfrak{q})}-E_L^{(1)}$ for 
$\mathfrak{q}=2,3,\cdots, 7$ correspond to
blue, red, green, purple, orange, and light blue
dots, respectively. 
The $n$th set of branches of $E_L^{(\mathfrak{q})}-E_L^{(1)}$ 
has at most $(n-1)$-fold quasi-degeneracy. 
}
\label{fig3}
\end{figure}

Several low-lying eigenvalues $E_L^{(\mathfrak{q})}$  
measured from the yrast-state energy $E_L^{(1)}$ 
are plotted in Fig.~\ref{fig3}.    
Each branch is specified by a particular set of particle-hole pattern, 
and is linear with respect to $L/N$. 

Let us calculate the first excitation energy 
from the yrast state $\Delta E_L \equiv E_L^{(2)} - E_L^{(1)}$. 
By evaluating the quasimomenta for $\mathfrak{q}=2$ 
as $\tilde{k}_j^{(2)}\equiv \tilde{k}[I_j^{(2)}]$,  
it is given by  
\be \label{DeltaEgen}
\Delta E_L = \sum_{\mathfrak{q}=1, 2} 
\left( \epsilon(\tilde{k}_p^{(\mathfrak{q})}) - \epsilon(\tilde{k}_h^{(\mathfrak{q})}) \right) 
 + O(1/\ell^2) . 
\ee
Here, the first term in Eq. (\ref{eq:finite-energy}) 
changes for the states of $\mathfrak{q}=1, 2$ 
only by the order $O(1/\ell^3)$ 
due to Eq. (\ref{eq:zero}).

The excitations from $\mathfrak{q}=1 \to 2$ are 
constructed either of the following processes, as shown in Fig.~\ref{fig2}.  

(a) Right excitation: move a particle at $I_p^{(1)}$ to $I_p^{(1)}+1$, and 
move a hole at $I_h^{(1)}$ to $I_h^{(1)}+1$. 
As a result, two particles at $\{I_p^{(2)}\}=\{I_p^{(1)}+1,\ I_h^{(1)}\}$ and two holes 
at $\{I_h^{(2)}\}=\{I_p^{(1)},\ I_h^{(1)}+1\}$ are excited relative to the yrast state. 
We denote the excitation energy associated with these ``right" processes 
as $\Delta E_L^{+}$. 
We have  
\bea\label{eq:Ep}
&& \Delta E_L^{+} =
\epsilon(\tilde{k}^{(2)}[I^{+}+2]) -  \epsilon(\tilde{k}^{(1)}[I^{+}+1]) \nonumber\\
& &\quad + \epsilon(\tilde{k}^{(1)}[I^{+}+1-L]) -  \epsilon(\tilde{k}^{(2)}[I^++2-L]). 
\eea 
In the TG limit, Eq.~(\ref{eq:Ep}) is simplified to 
$\Delta E_L^+/A =2L$ where $A=({\hbar^2}/{2m}) ({2\pi}/{\ell})^2$. 

(b) Left excitation: move a particle at $I^-$ to $I^--1$, and 
move a hole at $I_h^{(1)}$ to $I_h^{(1)}-1$.  
We thus have particles at $\{I_p^{(2)}\}=\{I^{-}-1, \ I_h^{(1)}\}$ and holes at $\{I_h^{(2)}\}=\{I^-,\ I_h^{(1)}-1\}$. 
The left excitation energy reads, 
\bea\label{eq:Em}
& & \Delta E_L^{-} =
\epsilon(\tilde{k}^{(2)}[I^{-}-1]) -  \epsilon(\tilde{k}^{(1)}[I^{-}]) \nonumber\\
&& \quad + \epsilon(\tilde{k}^{(1)}[I^{+}+1-L]) -  \epsilon(\tilde{k}^{(2)}[I^{+}-L]), 
\eea
which reduces to $\Delta E_L^-/A = 2(N-L)$ in the TG limit.

The first excitation energy $\Delta E_L$ as a function of $L$ is given by  
\bea \label{eq:DeltaE}
\Delta E_L= \left\{ \begin{array}{ll}  
\Delta E_L^- & :\quad L=0,1 , \\
\Delta E_L^+ & :\quad 2 \le L \le N/2 , \\
\Delta E_L^- & :\quad N/2 \le L \le N-2 , \\
\Delta E_L^+ & :\quad L=N-1, N , 
\end{array} \right.
\eea
where we regard the absence of a particle at $I^++1$ in the ground state as a hole at $I_h^{(1)}$ for $L=0$. 
Similarly for $L=N$, we regard the absence of a particle at $I^-$ in the ground state as a hole at $I_h^{(1)}$. 
For $L=1$ ($L=N-1$) it is obvious from Fig.~\ref{fig2} 
that the right (left) 
excitation is impossible 
because there is only one particle next to the hole at $I^{(1)}_h$. 
The $\Delta E_L$ is symmetric with respect to $L=N/2$, and 
%takes the minimum value 
minimal at $L=2$ and $L=N-2$.

Through the asymptotic expansion with respect to $\ell$, 
the left and right excitation energies $\Delta E_L^{\pm}$ 
are given in terms of the ground-state quasimomenta as follows. 
\be 
\Delta E_L^{\pm} = {\frac 1 {\ell}} \left( \mp  
 {\frac  {\epsilon^{'}(k[I_h^{(1)}])} {\rho(k[I_h^{(1)}])}} 
+ {\frac  {\epsilon^{'}( Q_0 )} {\rho(Q_0)}} \right) \, 
+ O(1/\ell^2). \label{eq:DeltaE}
\ee 
The difference between $\tilde{k}_j^{(\mathfrak{q})}$ and $k_j$ 
is of the order $O(1/\ell)$ for each $j$. It is also the case 
for ${\tilde k}_h$ and ${k}_h$, and hence 
we have put the ground-state momentum 
${k}_h$ in Eq. (\ref{eq:DeltaE}).

Some branches with different particle-hole configurations are 
``quasi-degenerate'' as shown in Fig.~\ref{fig3}.  
 For the $n$th lowest set of $E_L^{(\mathfrak{q})}$s, 
at most $n-1$ energy levels 
are lying within the range of the order $O(1/\ell^2)$. 
Here, the energies of higher excited states 
$\mathfrak{q}\ge 3$ are systematically calculated 
by giving particle-hole configurations similarly to the case $\mathfrak{q}=2$. 
Moreover, for $L$ close to the integral multiple of $N$,  
particle-hole excitations cost large energy due to tight 
constraints on configurations.

In order to numerically implement the above method, 
we calculate $\rho(k)$ and $\epsilon(k)$ solving the integral equations~(\ref{eq:rho}) and (\ref{eq:epsilon}), 
respectively,  setting $Q^{\pm}=\pm Q_0$.  
Integrating $\rho(k)$ we obtain 
the counting function $Z(k)$ 
and the ground-state quasimomenta $k[I_j]$'s 
through the inverse as $k_j=Z^{-1}(I_j/\ell)$.  
With $k_j$'s put 
in Eq.~(\ref{eq:DeltaE}) 
we can evaluate $\Delta E_L$ at an arbitrary strength of interaction.

%%%%%%%%%%%%%%%%%%%%% figure %%%%%%%%%%%%%%%%%%%%%%%%%%%
%%%%%%%%%%%%%%%%%%%%%%%%%%%%%%%%%%%%%%%%%%%%%%%%%%%

%%%%%%%%%%%%%%%%%%%%%%%%%%%%%%%%%%%%%%%%%%%%%%%%%%%%%%%%%%%%%%%%%%%%%%%%%%%%%%%%%%%%%%%%%%%%%%%%%%%%%%%%%%

Finally we discuss the yrast state in the light of the CFT. 
The low-lying excitation spectrum 
$E^{(\mathfrak{q})}_{L^{'}, N^{'}}$ 
among such eigenstates with the same $I_p$ and $I_h$ is given by    
\be
E^{(\mathfrak{q})}_{L^{'}, N^{'}} - E^{(1)}_{L, N} = 
\frac{2\pi v_F }{\ell}
\left(
{\frac {(\Delta N)^2} {4 \xi_0^2}} + \xi_0^2 (\Delta D)^2 + n_+ + n_- \right). 
\ee
Here $\Delta D$ represents the umklapp process with  
 $\Delta D/\ell 
= ({\tilde Z}_{\ell}(Q^{+}) + {\tilde Z}_{\ell}(Q^{-}))/2$, positive integers 
$n_{\pm}$ denote particle-hole excitations near $Q^{\pm}$, 
and $\Delta N=N^{'}-N$. The Fermi velocity $v_F$ is given by 
$v_F= \epsilon^{'}(Q_0)/(2\pi \rho_{\ell}^{p,h}(Q_0))$.  
We define dressed charge $\xi(k)$  by  
$\xi(k) = 1 - \int_{-Q_0}^{Q_0}  
K(k-q) \xi(q) dq/{2\pi}$, and $\xi_0 = \xi (Q_0)$.

When $L$ is an integral multiple of $N$; i.e., when the system undergoes the center-of-mass rotation, 
the system is mapped to the $L=0$ state in a rotating frame by the Galilean transformation. 
In this case the low-lying excitation spectrum is  
consistent with those of the $U(1)$ CFT with central 
charge $c=1$.  The connection also holds   
for various integrable systems with finite angular momentum. 
It enables us to predict various scaling properties of correlation functions, 
such as scaling exponents.  
It should play a significant role in the study of spectral properties for rotating systems of interacting 1D bosons 
and in the connection to solitons \cite{IT80,KCU10}.

In summary, we have constructed excited states with a fixed angular momentum and evaluated the excitation energies in terms of dressed energy through the finite-size asymptotic expansion. These energies include a finite-size correction 
of the order $O(1/\ell)$, which would be detected 
in the ultracold atomic gases. 
We have found that higher yrast excited states have 
quasi-degenerate fine structures. 
The special classes of angular momentum states, 
where $L$ is given by an integral multiple of $N$, are related to  
the $U(1)$ CFT. 
This relevance will provide a powerful scheme 
for the analysis of the spectra, 
correlations, and scattering experiments.

The authors would like to thank F.~G{\"o}hmann
 and M.~Ueda for their useful discussions. 
The present research is partially supported by 
Grant-in-Aid for Scientific Research No. 20540365 and No. 21710098. 
J. S. is supported by Grant-in-Aid for JSPS fellows. 

% _______________________________________________________________________________

\end{document}